\documentclass[12pt]{article}

\RequirePackage{amssymb}
\begin{document}
\begin{center}
{\bf  Physical Framework of Quantization Problem\footnote{
First published by the Seminar of Geometry and Topology from the 
University of Timisoara, Romania, as preprint 102/1990.} } \\[1cm]
Marius Grigorescu \\[1cm] 
\end{center}
\noindent
{\bf Abstract}: The paper presents shortly the geometric 
approach to the problem of a general quantization formalism, both 
physically meaningful and mathematically consistent. 

The notion of quantization has appeared at the beginning of the century in the theory of
heat radiation, since M. Planck has formulated the hypothesis of the energy quanta 
\cite{mpl}.
This hypothesis assumed a finite number of ways to distribute the energy of the
$\nu$-frequency radiation over a given number of oscillators, every distribution assigning 
to each oscillator integer multiples of the "energy quantum" $\epsilon = h \nu$,
$h=6.626 \times 10^{-34}$ J$\cdot$s. Later on, the Planck's hypothesis was interpreted
as a consequence of a "quantum constraint" acting in general on the harmonic oscillator
considered as classical system. Thus, the problem of quantization was to "select an 
infinite, discrete number of quantum possible real motions, from the continuous
manifold of all mechanically possible motions" \cite{som}, by an appropriate system of constraints.
The action of these constraints on the classical motion as a whole, and not on the 
physical state at a given time, has suggested to express them as restrictions on the
geometrical quantities associated with the trajectory in the space of the classical states.
These states were identified with the points of the phase space, locally parameterized
by generalized coordinates and momenta. The dimensional equality between the constant $h$
and the classical action has allowed to select the quantum trajectories for the separable 
systems with multiple periodic motions by using the conditions of Bohr, Wilson and
Sommerfeld (BWS), requiring the phase integrals $J_k = \oint p_k dq^k$, $k=1,2,...n$,
($n$ is the number of degrees of freedom ) to be integer multiples of the "action" quantum $h$:
\begin{equation}
J_k= \oint p_k dq^k = n_k h,~~n_k=0,1,2,... \label{1}
\end{equation}
Because $p_k$ is given by the solution $S(q^1,q^2,...,q^n, \alpha_1, \alpha_2, ...,
 \alpha_n)$ of the Hamilton-Jacobi equation:
\begin{equation}
H( \frac{ \partial S}{ \partial q} , q) = \alpha_1~~, \label{2}
\end{equation}
$\frac{ \partial S}{ \partial q} \equiv \{ \frac{ \partial S}{ \partial q^1}, ...,
\frac{ \partial S}{ \partial q^n} \}$, $q \equiv \{q^1,q^2,...,q^n \}$, as $p_k = 
\frac{ \partial S}{ \partial q^k}$, the condition (\ref{1}) selects a discrete set of
integration constants $\{ \alpha_1, \alpha_2, ..., \alpha_n \}$ corresponding to the
quantum allowed motions. \\ \indent
The application of these quantization formulas has provided a description in agreement 
with experiment for the energy levels of the hydrogen atom, or for the Stark and
Zeeman effects. Also, by quantizing the relativistic Kepler problem, the fine 
structure of the spectral lines was explained. \\ \indent
Beside its success, the method of phase integrals quantization was faced with
various shortcomings, and in particular it was not able to give a quantum description of
the free micro-particles. In 1924 Louis de Broglie has formulated the principle of
duality, extending the double nature of light, as wave and particle, to all forms of
matter. So, to a free micro-particle without spin, having the energy $E=mc^2$, $m=m_0/
\sqrt{1-v^2/c^2} $, $m_0=$ the rest mass, a scalar plane wave having the frequency 
$\nu = E/h$, the wavelength  $\lambda = h/mv$ and the group velocity 
$v_g = \frac{ d \omega}{dk}= \frac{ dE}{d(mv)} = v $, was associated. \\ \indent
The wave associated to the micro-particles placed in potential force fields was obtained
by extending the correspondence between the geometrical and wave optics. Within the
geometrical optics, the way of the light rays may be equivalently described by the wave
surfaces $S(x,y,z)=$ constant, orthogonal to the rays. The equation determining the 
function $S$ is formally identical with the Hamilton-Jacobi equation, and it can be 
obtained from the wave equation written for monochromatic waves in the limit of small
wave-lengths. This correspondence, together with de Broglie's formulas for the free
particle, has suggested a "kinematic" quantization, by expressing the refraction index
from the monochromatic wave equation through the classical potential energy. The final
equation obtained for the wave function $\Psi$, written as
\begin{equation}
\Delta \Psi + \frac{ 2m}{ \hbar^2} (E-V) \Psi =0 ~~,~~ \hbar= \frac{h}{ 2 \pi}~, \label{3}
\end{equation}
with $E$ and $V$ the total and the potential energy, respectively, has accurately solved
the problem of the energy spectra for all the potentials known as physically relevant.
\\ \indent
Intending to establish a dynamical equation, able to describe the time-evolution of the
micro-particles, the equation (\ref{3}) was generalized, but the result was proved to be
a breakdown with the classical wave picture. Thus, by contrast to the classical equation
for real waves, the fundamental quantum dynamical equation:
\begin{equation}
\hat{H} \Psi = i \hbar \frac{ \partial \Psi}{ \partial t} \label{4}
\end{equation}
contains only the first order time-derivative of $\Psi$. This equation can be obtained by
acting on $\Psi$ with the operator equation resulting when the derivatives 
$p= \frac{ \partial S}{ \partial q}$, $s= - \frac{ \partial S}{ \partial t}$ from the
time-dependent Hamilton-Jacobi equation $H(p,q,t)=s$, are replaced by the operators
$\hat{p} = - i \hbar \frac{ \partial}{ \partial q}$, $\hat{s} =  i \hbar \frac{
\partial}{ \partial t}$, but in general the non-commutativity of  $\hat{p}$
and $q$ leads to ambiguities in the definition of the operator $\hat{H}$.
Consequently, the further line of reasoning was to take the dynamics of $\Psi$ as
basically independent on a underlying classical picture, and to see the correspondence
between the operator $\hat{H}$ and the classical Hamilton function $H(p,q,t)$, if any,
merely as an accident. Within the new formulation, known as quantum mechanics\footnote{
Developed as "matrix", respectively "wave" mechanics, by  W. Heisenberg (July, 1925) 
and E. Schr\"odinger (1926).}, the function $\Psi$ is not an observable, 
and moreover, it is not a wave in the usual three-dimensional space, but it is a 
representative element for the "quantum state" of the system. This state is a ray in an 
abstract Hilbert space ${\cal H}$, without any classical correspondent, associated to the 
quantum system. Its dynamics is determined by the time-dependent Schr\"odinger equation (\ref{4}), 
where $\hat{H}$ is a self-adjoint operator defined on a dense domain in ${\cal H}$. Because 
the rays of ${\cal H}$ and the operators on ${\cal H}$ are not observables, the "quantization 
problem" was apparently eliminated, but in fact it was shifted towards the connection with 
experiment. Before analyzing more closely this problem, it is interesting to remark
that even the fundamental quantum equation (\ref{4}) contains a variable considered to be
classically defined, namely the time $t$. More generally, the whole space-time
framework is classical, and all the observables are classical quantities, subject to
classical dynamics. \\ \indent
Defining the value of an observable in a given state $\Psi$ as the mean-value 
$\langle \Psi \vert \hat{A} \vert \Psi \rangle$ of a symmetric operator $\hat{A}$
which is associated to the observable, the kinematic task of the quantization problem
was to construct the Hilbert space ${\cal H}$ and the correspondence between 
observables and operators. To obtain this correspondence, Dirac (1930) has suggested
an algebraic method based on the construction of an isomorphism between the Lie
algebra of the operators on ${\cal H}$ and the Poisson algebra of the observables \cite{dir}.
Considering the set of observables be represented by the set ${\cal F}(M)$
of the smooth real functions over a classical phase space $(M, \omega)$, with 
$\omega$ the globally defined symplectic form, then it becomes a Lie algebra 
defining the Poisson bracket $\{ \cdot , \cdot \}$ by:
\begin{equation}
\{ f,g \} = \omega(X_f,X_g) =- L_{X_f} g,~~f,g \in {\cal F}(M)~~. \label{5}
\end{equation}
Here $X_f$ is the vector field determined by $i_{X_f} \omega = df$, and 
$L_{X_f}$ denotes the Lie derivative with respect to $X_f$. In the case when
$M=T^*Q$, $Q=R^n$, the symplectic form is $\omega_0 = \sum_{k=1}^n dq^k \wedge dp_k$, 
and the complete quantization of $Q$ was defined as an 
R-linear map $f \rightarrow \hat{f}$ from ${\cal F}(M) $ to a set 
${\cal A}({\cal H})$ of symmetric operators on the Hilbert space 
${\cal H}$, having the following properties \cite{am}: \\
1. the map $\hat{~}: {\cal F}(M) \rightarrow {\cal A}({\cal H})$ is injective. \\
2. $[\hat{f}, \hat{g} ] = i \hbar \hat{~} \{ f,g \}$, $f,g \in {\cal F}(M)$. \\
3. $\hat{1} = I$, where 1 is the unity function, constant on $M$, and $I$ is the
identity operator on ${\cal H}$. \\
4. $\hat{q}^k$, $\hat{p}_k$, $k=1,...,n$ act irreducibly on ${\cal H}$. \\
\indent
The first condition requires to have an associated quantum operator for each
observable. \\ \indent
The second condition may be taken as the consequence of the time-independence 
of the map $\hat{~}$ for a wide class of dynamical systems on $(M, \omega)$, and is
directly related to the dynamical problem of quantization. In particular, if the
observable $f \in {\cal F}(M)$ has a complete Hamiltonian field (the integral
curves $c_t$ through $c_0=m$ given by $\dot{c}_t = X_f(c_t)$ are defined for 
$\forall t \in ( - \infty, \infty )$, $\forall m \in M$), then the associated 
operator $\hat{f}$ must be self-adjoint, and generates a one-parameter group
of unitary transformations on ${\cal H}$, $U^t_{( \hat{f})} = \exp (- \frac{i}{
\hbar} \hat{f} t )$. If the map $g \rightarrow \hat{g}$, $g \in {\cal F}(M)$
is time-independent, then $( \hat{g})_t = \hat{~}(g_t)$, where $g_t= F^*_tg=
g \circ F_t$, ($F_t$ is the flow of $X_f$, $c_t=F_t(m)$), and 
$( \hat{g})_t =  (U^t_{( \hat{f})})^{-1} \hat{g} U^t_{( \hat{f})} $. Locally, the
equation $\hat{~}( g \circ F_t) =  (U^t_{( \hat{f})})^{-1}  \hat{g} U^t_{( \hat{f})}$
becomes $\hat{~}( L_{X_f} g) =  \frac{i}{ \hbar} [ \hat{f}, \hat{g} ]$, or
$\hat{~} \{ f,g \}  = - \frac{i}{ \hbar} [ \hat{f}, \hat{g} ]$. \\ \indent
The third condition is imposed by the construction of the observables, accounting 
for the uncertainty relations of Heisenberg, or for the empiric wave-particle
duality. Among the observables there are also constants which are not naturally
associated to the phase-space geometry, as the interaction strengths, the electric
charge, or the mass. These act multiplicatively on the state vectors, as any real 
constant, behaving as multiples of the identity operator on ${\cal H}$. The third
condition requires to quantize all the constants in the same way, irrespective if
these are coming from the Poisson bracket of some observables (as the canonically
conjugate coordinates), or if they represent dynamical constants of the classical
system. In particular, the Heisenberg uncertainty relations are a consequence of
this condition, proving its sufficiency. \\ \indent
The fourth condition accounts for the requirement of expressing the action of
every operator associated to an observable through the action of the operators
$(\hat{q}^k, \hat{p}_k)$, $k=1,...,n$. This requirement corresponds to the 
classical condition of expressing every observable as a function of the particular
observables represented by the canonical coordinates $(q^k,p_k)$, $k=1,...,n$ on
the phase space. Showing that every irreducible representation of the algebra
${\cal C}_0 \equiv (q^1,q^2,...,q^n,p_1,p_2,...,p_n,1)$  is unitarily equivalent to
the Schr\"odinger representation, where ${\cal H} = {\cal L}^2(R^n)$, $\hat{q}^k = q^k$,
$\hat{p}_k = - i \hbar \frac{ \partial}{ \partial q^k} $, the Stone - von Neumann
theorem (1932) \cite{am} has given explicitly the space ${\cal H}$ and the map
$f \rightarrow \hat{f}$. \\ \indent
In 1951 van Hove has proved rigorously the incompatibility between the four 
conditions stated above, proving therefore the impossibility of a complete
quantization \cite{am}. Moreover, he has shown that it is possible to fulfill the
first three conditions, obtaining a "prequantization", but then the algebra
${\cal C}_0$ is represented with infinite multiplicity. Also, if only the last
three conditions are imposed, then the application $\hat{~}$ must be restricted to
some subalgebra ${\cal C} \subset {\cal F}(M)$, containing ${\cal C}_0$. In
particular, the polynomial Hamiltonians quantizable within the Schr\"odinger
representation are at most of second degree in coordinates and momenta.\\ \indent
These results have shown that the problem of the algebraic quantization has no 
meaning for the set of all observables contained in ${\cal F}(M)$, but only for 
a couple of subalgebras, $({\cal C}_0(M), {\cal C}(M) )$, ${\cal C}_0(M) \subset
 {\cal C}(M) \subset {\cal F}(M)$, with ${\cal C}_0$ represented irreducibly. 
\\ \indent
For the manifolds $(M, \omega)$ which are not diffeomorphic to $(T^*R^n, \omega_0)$,
(the de Rham cohomology class $[ \omega ]_{dR} \ne 0$), the canonical coordinates
$(q^k,p_k)$, $k=1,2,...,n$ are only locally defined and are not observables, so that
the fourth condition must be reformulated. A direct generalization concerns the
homogeneous symplectic spaces $(M, \omega)$ onto which a Lie group $G$ acts transitively
and strongly symplectic. The action of $G$ is defined globally by a homomorphism 
$\sigma : G \rightarrow Ham(M)$ and locally by the map $d \sigma : g \rightarrow ham(M)$.
Here $Ham(M)$ is the group of Hamiltonian diffeomorphisms of $M$, $g$ denotes the Lie
algebra of $G$, and $ham(M)$ is the Lie algebra of Hamiltonian fields on $M$. A lift of
the map $d \sigma$ is a Lie algebra homomorphism $\lambda : g \rightarrow {\cal F}(M)$ 
\cite{gs} such that the following diagram is commutative.
\\ 
\begin{center} 
\( \begin{array}{ccc}
 0 ~ \rightarrow ~R ~  \rightarrow & {\cal F} (M) ~ \rightarrow  ~ ham(M)  &  \rightarrow ~ 0 \\ 
~~~~~~~ & ~~~~~_{\lambda}  \nwarrow ~~  \uparrow_{d \sigma} & \\ 
~~~~~~~ & ~~~~~~~~ g &  
\end{array} \) 
\\ 
\end{center} 
This lift exists iff the 2-cocycle $\mu : g \times g \rightarrow R$,
\begin{equation}
\mu ( x,y) = \{ \lambda_0(x), \lambda_0(y) \} - \lambda_0( [x,y])~,~~x,y \in g \label{6}
\end{equation}
(with $\lambda_0 : g \rightarrow {\cal F}(M)$ any linear map making the diagram to
commute), has a vanishing de Rham class $[ \mu]_{dR} \in H^2(g,R) $ \cite{gs, kos}. 
Using the map $\lambda$, the fourth condition defining the Hilbert space can be
formulated as: \\
4'. If $(M, \omega)$ is a strong symplectic homogeneous $G$-space, and $\lambda$ is a
lift of the map $d \sigma$, then the subalgebra ${\cal C}$ of the quantizable
observables must contain the subspace $\lambda(g)$, and the representation Hilbert
space ${\cal H}$ must be irreducible with respect to the action of the operators from
$\hat{~}( \lambda(g) )$ \cite{woo}. \\ \indent
As a first step towards quantization, the prequantization of the algebra ${\cal F}(M)$
was obtained geometrically \cite{kos, woo, sou} by constructing a Hilbert space 
${\cal H}_M$ as a space $\Gamma_L$ of sections in a hermitian line bundle with connection 
$(L, \alpha)$ over $(M, \omega)$.  Requiring the symplectic form $\omega$ to be the
curvature of the connection \cite{kos} ($d \pi^* \alpha = \omega$, with $\pi : L 
\rightarrow M$ the projection on the base in $L$), the line bundle exists only in special
cases, namely when $\omega$ fulfills the Weil's integrality condition: $h^{-1} [ \omega 
]_{dR} \in Z$. This condition acts similarly to the BWS condition of the old quantum
mechanics, selecting the classical manifolds of observables compatible with the 
description of a quantum system. In particular, for the integrable systems the Weil condition
for the reduced phase space and BWS quantization rule become the same. \\ \indent
In the case $M=T^*R^n$ the prequantum Hilbert space ${\cal H}_M$ is represented by:
\begin{equation}
{\cal L}^2 (R^{2n})= \{ \Psi:M \rightarrow C, \int \frac{\vert \omega^n \vert}{n!} \vert \Psi \vert^2 
< \infty \} \label{7}
\end{equation}
where $\vert \omega^n \vert/ n!$ is the natural volume on $R^{2n}$. Because this space is too
large for quantization, it was investigated the possibility to extract a "quantum subspace"  
${\cal H}_P \subset {\cal H}_M$ generated by the sections from $\Gamma_L$ which are constant on
the leaves of an integrable foliation of $M$ by Lagrangian submanifolds (a polarization $P$
on $M$). The subspace ${\cal H}_P$ is not invariant to the action of all the operators 
associated with observables, and consequently the subalgebra of the observables quantizable 
within ${\cal H}_P$ is ${\cal C}_P = \{ f \in {\cal F}(M), \hat{f} {\cal H}_P \subset 
{\cal H}_P \}$. \\ \indent
This procedure of quantization has the advantage of giving a natural correspondence between
the spaces ${\cal H}_P$ and ${\cal H}_{P'}$ associated to different polarizations $P$ and
$P'$, as subspaces of the whole prequantum space ${\cal H}_M$. Unfortunately, it cannot be
physically accepted because in many important cases, including the Schr\"odinger 
quantization of $T^*R^n$, ${\cal H}_P$ does not exist. \\ \indent
An alternative solution is to construct an intrinsic Hilbert space ${\cal H}_P^{[ \mu]}$
associated to the polarization $P$ independently of the prequantum space ${\cal H}_M$.
With some enough general restrictions, the space ${\cal H}_P^{[ \mu]}$ corresponds to a
class of measures $[ \mu]$, Lebesgue equivalent, defined on a Lagrangian submanifold
$Q \subset M$, transversal to $P$, whose elements are called half-densities. This formalism
has solved the problem of quantization for a wide class of physically relevant Hamiltonians,
but the results are still slightly different from those obtained by using the standard quantum
mechanics, well supported by experiments. Thus, the energy $\epsilon_n = n h \nu$, $n=0,1,2,...$
predicted for the one-dimensional harmonic oscillator is different from the correct result
$\epsilon_n = (n+1/2) h \nu$ \cite{woo}. Such differences were further eliminated by assigning
a metalinear structure to the Lagrangian submanifold $Q$, and correspondingly extending the
half-densities Hilbert space ${\cal H}_P^{[ \mu]}$ to a Hilbert space of half-forms 
${\cal H}_P^{\Lambda_{1/2}}$ \cite{woo}. As a rule, its construction is cumbersome, and the
half-forms quantization was tested only for a small number of simple systems. \\
 
\end{document}